\documentclass[12pt,a4paper]{iopart}
\usepackage{ae}
\usepackage{graphicx}
\usepackage{color}
\usepackage{wasysym}
\usepackage{subfigure}
\usepackage{siunitx}
\usepackage{xspace}
\usepackage{overpic}
\usepackage{float}
\usepackage[numbers,sort&compress]{natbib}
\usepackage[all]{nowidow} 
\usepackage{soul}

\usepackage{hyperref}
\hypersetup{
	colorlinks=true,
	pdfstartview=FitV,
	linkcolor=blue,
	citecolor=blue,
	urlcolor=blue,
	pdfauthor={Hoppe Mathias, hoppe@chalmers.se},
	plainpages=false,
}

\newcommand{\SOFT}{\textsc{SOFT}\xspace}

\newcommand{\CODE}{\textsc{CODE}}

\newcommand{\thetap}{\ensuremath{\theta_\mathrm{p}}\xspace}
\newcommand{\lambdac}{\lambda_\mathrm{c}}

\renewcommand{\d}{\mbox{d}}

\newcommand{\Eq}[1]{Eq.~(\ref{#1})}

\newcommand{\Fig}[1]{Fig.~\ref{#1}}

\newcommand{\eqref}[1]{(\ref{#1})}

\newcommand{\white}[1]{\textbf{\textcolor{white}{#1}}}

\begin{document}

\title[SOFT: A synthetic synchrotron diagnostic for runaway electrons]{SOFT: A synthetic synchrotron diagnostic for runaway electrons}

\author{
    M Hoppe$^{1}$,
    O Embr\'eus$^{1}$, 	
	RA Tinguely$^{2}$,
	RS Granetz$^{2}$,
	A Stahl$^{1}$, and
	T F\"ul\"op$^{1}$
}

\address{$^{1}$ Department of Physics, Chalmers University of Technology, G\"oteborg, Sweden}
\address{$^{2}$  Plasma Science and Fusion Center, Massachusetts Institute of Technology, Cambridge MA, USA}
\ead{hoppe@chalmers.se}

\begin{abstract}
Improved understanding of the dynamics of runaway electrons can be obtained by measurement and interpretation of their synchrotron radiation emission.  Models for synchrotron radiation emitted by relativistic electrons are well established, but the question of how various geometric effects -- such as magnetic field inhomogeneity and camera placement -- influence the synchrotron measurements and their interpretation remains open. In this paper we address this issue by simulating synchrotron images and spectra using the new synthetic synchrotron diagnostic tool \SOFT~(\emph{Synchrotron-detecting Orbit Following Toolkit}). We identify the key parameters influencing the synchrotron radiation spot and present scans in those parameters. Using a runaway electron distribution function obtained by Fokker-Planck simulations for parameters from an Alcator C-Mod discharge, we demonstrate that the corresponding synchrotron image is well-reproduced by \SOFT~simulations, and  we explain how it can be understood in terms of the parameter scans. Geometric effects are shown to significantly influence the synchrotron spectrum, and we show that inherent inconsistencies in a simple emission model (i.e.~not modeling detection) can lead to incorrect interpretation of the images.
\end{abstract}

\section{Introduction}
Runaway acceleration of charged particles is one of the most interesting phenomena in plasmas. If the electric field exceeds a critical value, a fraction of particles can be detached from the bulk population and accelerated to relativistic energies \cite{Dreicer1959}.  Electrons are usually detached most easily and are then referred to as {\em runaway  electrons}. In magnetic-fusion plasmas, runaway electrons can be generated during disruption events, in which a rapid cooling of the plasma takes place and a large electric field is induced in the toroidal direction in order to maintain the plasma current (which can be several megaamperes). A runaway-electron beam containing particles with energies of up to tens of MeV, carrying a significant fraction of the initial current, may form in such situations. If the density is sufficiently low, such a beam may also form during start-up or flat-top operation. The potential for damage by such a localised beam of highly-energetic particles upon contact with the vessel wall is substantial, particularly in view of future large tokamaks, such as ITER, since the runaway generation rate increases exponentially \cite{RosenbluthPutvinski} with the plasma current. This damage must be avoided or reduced using mitigation techniques \cite{Hollmann2015} to make stable and reliable ITER operation possible. To this end, many dedicated experiments are performed on a number of existing tokamaks to study the dynamics of relativistic runaway-electron beams in both quiescent and disruptive plasmas.

One of the primary methods to experimentally diagnose runaway beams is to study the synchrotron radiation they emit. Synchrotron radiation is emitted by relativistic electrons as a consequence of their gyro motion and is an attractive means of diagnosing the runaways since it does not require them to leave the plasma. In this paper we analyze the effects of geometry and intensity distribution of runaway synchrotron radiation by modeling synchrotron radiation images, which in typical runaway scenarios are experimentally observed in the visible and infra-red spectral ranges. The pioneering measurements of synchrotron radiation from runaway electrons were made already in the 1990's \cite{Finken1990,Jaspers1995} and since then such measurements have become a routine diagnostic used in many tokamaks around the world, including TEXTOR \cite{Jaspers2001,Wongrach2014}, FTU \cite{Esposito2003,Esposito2017}, Alcator C-Mod \cite{Tinguely2015}, ASDEX Upgrade \cite{Papp2016IAEA}, TCV \cite{Papp2016IAEA}, COMPASS \cite{Vlainic2015} DIII-D \cite{Yu2013}, EAST \cite{Zhou2014}, and KSTAR \cite{Cheon2016}.

Although many observations of synchrotron radiation emitted by runaway electrons exist, obtaining useful information about the spatial and velocity distributions of the runaways requires careful interpretation of the diagnostics, which provide primarily line- and volume-integrated information (i.e.~a camera photographing the plasma). In particular, modeling the synchrotron emission as coming from particles with a single energy and pitch angle is insufficient and can be misleading. Further complications are introduced by the geometry of the problem (in particular the twist of the magnetic field lines and the relativistic forward beaming of the synchrotron radiation); the pitch-angle, energy and radial distributions of runaway electrons; and the position and sensitivity of the camera. Theoretical descriptions of the relation between runaway and plasma parameters and the extent of the observed radiation spot have been considered previously \cite{Pankratov1996,EntropSnakes1999,Zhou2014,Gomez2017}, however, with the exception of~\cite{Gomez2017}, these studies did not consider the synchrotron intensity distribution in the images (only the overall spot shape), and did not include effects of the velocity-space distribution of the runaway population.

Recently, a new synthetic synchrotron diagnostic taking full-orbit effects into account was presented in Ref.~\cite{Gomez2017}, where synchrotron images and spectra were also calculated and discussed with respect to the full-orbit effects. In contrast to Ref.~\cite{Gomez2017}, we derive a rigorous synthetic diagnostic theory expressed in terms of guiding-center quantities to reduce the numerical complexity of the problem. Using the simplified model we also analyze and discuss synchrotron images in terms of properties of the distribution function and detector.

The aim of the present paper is to analyze the synchrotron image and spectrum obtained from a population of relativistic electrons -- with arbitrary spatial and velocity-space distributions -- in an arbitrary tokamak magnetic geometry using the synthetic synchrotron diagnostic SOFT (Synchrotron-detecting Orbit Following Toolkit)~\cite{HoppeMSc,SOFT}. In Section~\ref{sec:soft}, the underlying theory used in the simulations is presented using a number of physically motivated approximations to reduce the computational cost of the problem. In Section~\ref{sec:comparisons}, a thorough investigation of the parameters that influence the final image is performed. The tool will also be used to analyze a specific image of synchrotron emission from a runaway beam, obtained in the Alcator C-Mod tokamak (Section~\ref{sec:Alcator_comparison}).

\section{Synthetic synchrotron diagnostics}
\label{sec:soft}
\noindent
The basic idea of simulating the synchrotron radiation from runaway electrons is simple: place a detector in the tokamak, follow a large number of particle trajectories in space, and at each point of space compute their contribution to the synchrotron image based on the particles' velocities at that point. In this section we will discuss the mathematical theory of this scheme. The angular and spectral distributions of synchrotron radiation will be introduced and discussed briefly, followed by a discussion of how a synchrotron image can be interpreted. In the discussion we will consider an ideal detector, i.e.~we will assume that the detector has a uniform response function in a given spectral interval and directly measures the electromagnetic energy flux reaching the aperture.

\subsection{Theory of synchrotron detectors}\label{sec:synthdiagtheory}
The radiated power detected at time $t$ in the wavelength interval between $\lambda$ and $\lambda+\d\lambda$ per unit surface area $\d A$ of a detector, per unit solid angle $\d\Omega_{\mathbf{n}}$ centered on the vector $\mathbf{n}$ by the observer, located at $\mathbf{x}_0$, is
\begin{eqnarray}\label{eq:softfund}
    \hspace{-20mm}\frac{\d^2 P_0(\mathbf{x}_0,\,\mathbf{n},\,\lambda,\,t)}{\d \lambda \d A\d\Omega_{\mathbf{n}}} &= \int \delta^2\left(\mathbf{r}/r-\mathbf{n}\right) \frac{\mathbf{n}\cdot\hat{\mathbf{n}}}{r^2} \frac{\d^2 P(\mathbf{x},\,\mathbf{p},\,\mathbf{x}_0,\,\lambda)}{\d \lambda \d\Omega} f(\mathbf{x},\,\mathbf{p},t-r/c) \, \d\mathbf{x}\d\mathbf{p},
\end{eqnarray}
where $f$ is the particle distribution function, $\mathbf{x}$ is the particle's position, $\mathbf{p} = m\mathbf{v}/\sqrt{1-\beta^2}$ is the momentum, $m$ is the particle's rest mass, $\beta = |\mathbf{v}|/c$ is the particle's speed normalized to the speed of light in vacuum $c$, $\d^2 P / \d\lambda \d\Omega$ is the emitted power per unit wavelength interval and solid angle~\cite{Blumenthal1970,Ginzburg1968}, $\mathbf{r} = \mathbf{x}-\mathbf{x}_0$ is the relative position between particle and detector, $\thetap$ is the particle's pitch angle, $\mu$ is the angle between the magnetic field and $\mathbf{n}$ and $\hat{\mathbf{n}}$ is the direction in which the detector is facing (i.e.~the unit normal vector of the camera lens). The delta function singles out the radiation travelling along the specified line-of-sight $\mathbf{n}$, and $\d\Omega = \frac{\mathbf{n}\cdot\hat{\mathbf{n}}}{r^2}\d A$ relates the solid angle element to detector surface area element.

Equation~\eqref{eq:softfund} describes the radiation reaching a point on the detector along a direction $\mathbf{n}$ at a time $t$ from all points in phase-space. The contribution to an individual pixel of the 2D grid of pixels making up the synchrotron image, which we label by the indices $(i, j)$, in a small wavelength interval $\d\lambda$, is obtained by integrating over the finite surface of the detector and all lines-of-sight within the region $\mathbf{N}_{ij}$ corresponding to pixel $(i, j)$:
\begin{equation}\label{eq:softfull}
    \frac{\d I_{ij}}{\d\lambda}(\mathbf{x}_0, t) = \int_{A} \int_{\mathbf{N}_{ij}} \frac{\d^2 P_0(\mathbf{x}_0,\,\mathbf{n},\,\lambda,\,t)}{\d \lambda \d A} \d A \d\mathbf{n},
\end{equation}
where $A$ denotes the detector surface.

According to Liouville's theorem, the distribution function is constant along particle orbits when the system evolves slowly in time compared to the orbital time scale. Therefore, the distribution need only be specified at a single point along each orbit, from which it can be determined at all other points by integrating the equations of motion. By utilizing a particular orbit-coordinate set which we describe below (similar to the set introduced in \cite{RomePeng1979}), we can describe the radiation from any distribution function in terms of a reduced three-dimensional phase space.

The first coordinate transformation is a standard, zeroth-order, guiding-center transformation~\cite{Northrop1961,Tao2007}, as it is numerically more efficient to solve for the guiding-center orbit rather than a full particle orbit. We let $\mathbf{X}$ denote the guiding-center position, $p_\parallel$ the particle momentum parallel to the magnetic field, $p_\perp$ the magnitude of the particle momentum perpendicular to the magnetic field, and $\zeta$ the gyrophase of the particle. To zeroth order, the Jacobian for the spatial transformation is unity while for the momentum transformation it is just $p_\perp$, and we get
\begin{equation}\label{eq:gcdiffel}
    \d\mathbf{x}\d\mathbf{p}\approx p_\perp\,\d\mathbf{X}\d p_\parallel \d p_\perp \d \zeta = |J_p| p_\perp^{(0)} \d\mathbf{X}\d p_\parallel^{(0)} \d p_\perp^{(0)} \d \zeta,
\end{equation}
where $\mathbf{p}^{(0)}$ is the particle momentum evaluated at any point $\mathbf{X}^{(0)}$ along its orbit, and where the conservation of magnetic moment $\mu = p_\perp^2/2m_e B$ yields a Jacobian
\begin{equation}
J_p = \frac{B(\mathbf{X})}{B(\mathbf{X}^{(0)})}\frac{p_\parallel^{(0)} }{\sqrt{\left(p^{(0)}\right)^2-\left(p_\perp^{(0)}\right)^2 \frac{B(\mathbf{X})}{B(\mathbf{X}^{(0)})}}} .
\end{equation}
The use of guiding-center coordinates simplifies the computation since the integral over $\zeta$ can be carried out explicitly, putting the integrand in~\Eq{eq:softfund} in a gyro-averaged form. In the limit of small gyro-radius, the vector $\mathbf{r}$ will be independent of $\zeta$, allowing the delta function and geometrical factor $\mathbf{n}\cdot\hat{\mathbf{n}}/r^2$ to be taken out of the $\zeta$-integral. Because of gyrotropy, the distribution function will be independent of the gyrophase $\zeta$ and so can also be moved out of the $\zeta$-integral. The only factor for which a gyrophase dependence must remain is in the spectral power $\d^2P/\d\lambda\d\Omega$, and as we will see later, the synchrotron power formulas can be analytically averaged over a gyro-period, allowing us to completely eliminate the gyrophase $\zeta$ from the integral.

Secondly, we make a standard cylindrical coordinate transformation, $(R, Z, \phi)$, with $R$ the radial distance to the particle from the axis-of-symmetry, $Z$ the vertical offset of the particle from the midplane and $\phi$ the toroidal angle. The spatial differential element $\d\mathbf{X}$ of~\eqref{eq:gcdiffel} then becomes
\begin{equation}
    \d\mathbf{X} = R\,\d R \d Z \d \phi.
\end{equation}

So far we have only expressed the integral in terms of guiding-center coordinates, but we have still not fully taken the geometry of the magnetic field into account. Since the tokamak is an axisymmetric device, the equations of motion determine the guiding-center distribution in the poloidal plane, and because of toroidal symmetry the distribution will be identical in every such plane. In the third and final coordinate transformation we therefore change the poloidal guiding-center coordinates $R$ and $Z$ to the poloidal ``trajectory coordinates'' $\rho$ and $\tau$. The coordinate $\rho$ is the major radius at which the guiding-center is located at the start of its orbit (i.e.~in the outer midplane), while $\tau$ is the ``orbit time'', defined such that $\tau = 0$ corresponds to the outer midplane and $\tau = \tau_0$ the point in the poloidal plane which it would take a time $\tau_0$ for the guiding-center to reach, if starting in the midplane. The coordinate $\tau$ can also be thought of as an alternative poloidal angle that takes the magnetic field geometry into account. The determinant of the Jacobian for this transformation depends on the particular magnetic field used, and to allow for any (numerical) magnetic field to be used we will write the differential element
as 
\begin{equation}
    \d R \d Z = |J|\d\rho\d\tau,\qquad J = \frac{\partial R}{\partial \rho}\frac{\partial Z}{\partial \tau} - \frac{\partial R}{\partial \tau}\frac{\partial Z}{\partial \rho}.
\end{equation}

Equation~\eqref{eq:softfund} can now be cast into its final form,
\begin{eqnarray}
    \frac{\d^2P_0(\mathbf{x}_0, \mathbf{n}, \lambda, t)}{\d\lambda\d A} = \int \d\rho \,\d p_\parallel^{(0)}\d p_\perp^{(0)}\,p_\perp^{(0)} f_\mathrm{gc}(\rho, p_\parallel^{(0)}, p_\perp^{(0)}, t)\times\nonumber\\
    \times\d\phi \d\tau |J_p||J| R(\rho,\tau) \frac{\mathbf{n}\cdot\hat{\mathbf{n}}}{r^2} \delta^2\left(\mathbf{r}/r - \mathbf{n}\right) \left\langle\frac{\d^2 P(\mathbf{X}, p_\parallel, p_\perp, \zeta, \lambda)}{\d\lambda\d\Omega}\right\rangle,\label{eq:soft}
\end{eqnarray}
where the distribution of guiding-centers $f_\mathrm{gc}(\mathbf{X}, p_\parallel, p_\perp, t) \approx 2\pi f(\mathbf{x}, p_\parallel, p_\perp, t-r/c)$, $\mathbf{p}^{(0)}$ denotes the guiding-center momenta at $\tau=0$ and the gyro-average $\langle\ldots\rangle$ of a quantity $F = F(\zeta)$ is defined as
\begin{equation}\label{eq:gyroaverage}
    \langle F\rangle = \frac{1}{2\pi}\int_0^{2\pi} F(\zeta)\d\zeta.
\end{equation}

Given the distribution of guiding-centers $f_{\mathrm{gc}}$ in the outer midplane, i.e.~at $\tau = 0$, the integral over $\tau$ in~\Eq{eq:soft} can be evaluated by solving the guiding-center equations of motion as an initial value problem in a set of points $\{\tau_n\}$, in order to obtain $\mathbf{X}(\rho, \tau, \phi)$, $p_\parallel(\rho, \tau, \phi)$ and $p_\perp(\rho, \tau, \phi)$ at those points. Standard methods for solving the guiding-center equations of motion can therefore be applied -- \SOFT~uses the familiar RKF45 scheme~\cite{Ackleh2009} -- making the numerical solver very flexible when it comes to taking the magnetic field geometry of the simulated device into account.

A feature of Eq.~\eqref{eq:soft} that can simplify computations is the fact that the integrand can be separated into one part consisting of the distribution function, and one part representing various geometric and emission effects, which we denote by $\hat{I}_{ij}(\rho,p_\parallel^{(0)},p_\perp^{(0)})$:
\begin{equation}
    I_{ij} = \int\d \rho\d p_\parallel^{(0)}\d p_\perp^{(0)} f_{\mathrm{gc}}(\rho,p_\parallel^{(0)},p_\perp^{(0)}) \hat{I}_{ij}(\rho,p_\parallel^{(0)},p_\perp^{(0)}).
\end{equation}
By pre-computing and tabulating $\hat{I}_{ij}$, we may generate images or spectra by integrating over $(\rho, p_\parallel^{(0)}, p_\perp^{(0)})$, which reduces to a set of multiplications. For a particular detector setup and magnetic field configuration, $\hat{I}_{ij}$ can be generated and used to quickly produce synthetic synchrotron data.

The model in~\Eq{eq:soft} has the advantage of only assuming that the plasma is optically thin to the radiation, so while our studies are focused toward the synchrotron radiation emitted by runaway electrons, one could in principle use the same approach to simulate other types of radiation emitted by charged particle species in an axisymmetric plasma.

\subsection{Expressions for the angular and frequency distributions of the radiation power}
\label{sec:radiation_models}
Synchrotron radiation is emitted almost entirely in the particle's direction of motion, and the spectrum is practically a continuum in the IR or visible spectral ranges in scenarios relevant to magnetic fusion. The most general description of the radiation received by a detector from a relativistic electron in helical motion in a homogeneous magnetic field is~\cite{Schwinger1949,Bekefi}
\begin{eqnarray}
    \left\langle \frac{\d^2 P}{\d\lambda\d\Omega} \right\rangle &= \frac{9e^2\beta^2\gamma^{12}\omega_B^3}{256\pi^3\epsilon_0c^2\gamma_\parallel^2} \left(\frac{\lambda_\mathrm{c}}{\lambda}\right)^4 \left(\frac{1-\beta\cos\psi}{\beta\cos\psi}\right)^2 (1-\beta\cos\thetap\cos\mu) \nonumber\\
    &\times \left[ K_{2/3}^2(\xi) + \frac{(\beta/2)\cos\psi\sin^2\psi}{1-\beta\cos\psi} K_{1/3}^2(\xi) \right].\label{eq:angspecdist}
\end{eqnarray}
Here, $\langle\ldots\rangle$ denotes a gyro-average as defined in~\Eq{eq:gyroaverage}, $e$ is the elementary charge, $\epsilon_0$ is the vacuum permittivity, $m_e$ is the electron rest mass, $\gamma = (1-\beta^2)^{-1/2}$ and $\gamma_\parallel = (1-\beta_\parallel^2)^{-1/2}$, where $\beta_\parallel$ is the electron speed parallel to the magnetic field normalized to the speed of light, $\omega_B = eB/\gamma m_e$ is the electron cyclotron frequency, and $B$ is the magnetic field strength. The angle $\psi$ is defined through the angle $\mu$ between the velocity vector of the guiding-center and the observer's line-of-sight as $\psi = \mu-\thetap$, where $\thetap = \arccos(v_\parallel / v)$ is the pitch angle. The functions $K_{1/3}(\xi)$ and $K_{2/3}(\xi)$ are modified Bessel functions, and their arguments are
\begin{equation}
    \xi = \gamma^3\frac{\lambdac}{\lambda}\sqrt{\frac{(1-\beta\cos\psi)^3}{(\beta/2)\cos\psi}},
\end{equation}
with $\lambdac = 4\pi m_ec\gamma_\parallel / 3\gamma^2 e B$ the critical wavelength, approximately corresponding to the peak of emission. The factor $1-\beta\cos\thetap\cos\mu$ compensates for the difference between the observed and actual particle distribution, due to the finite speed of light~\cite{Ginzburg1968} (effectively compensating for the apparent superluminal motion of relativistic particles approaching an observer, otherwise commonly associated with observation of relativistic jets from quasars~\cite{Pearson1981}).

Similar gyro-averaged expressions for the angular or spectral distributions of synchrotron radiation can be computed and are found to be~\cite{HoppeMSc}
\begin{equation}
    \hspace{-24mm}
    \left\langle \frac{\d P}{\d\Omega} \right\rangle = \frac{e^4B^2\beta_\perp^2 \gamma_\parallel^2}{16\pi^2\epsilon_0 \gamma^2 m_e^2 c} (1-\beta\cos\thetap\cos\mu) \left[ \frac{\kappa^3}{2}(3\eta^2 - 1) - \left(\frac{\sin\mu}{\gamma}\right)^2 \frac{\kappa^5}{8}(5\eta^2 - 1) \right],\label{eq:angdist}
\end{equation}
and~\cite{Schwinger1949}
\begin{equation}
    \left\langle \frac{\d P}{\d\lambda} \right\rangle = \frac{1}{\sqrt{3}}\frac{ce^2}{\epsilon_0\lambda^3\gamma^2} (1-\beta\cos\thetap\cos\mu) \int_{\lambdac/\lambda}^\infty K_{5/3}(l) \d l,\label{eq:specdist}
\end{equation}
where $\beta_\perp$ is the electron speed perpendicular to the magnetic field, and
\begin{eqnarray}
    \eta = \left[1 - \left(\frac{\beta\sin\mu\sin\thetap}{1 - \beta\cos\mu\cos\thetap}\right)^2 \right]^{-1/2},\\
    \kappa = \left\{(1-\beta\cos\psi)\left[1-\beta\cos\left(\psi+2\thetap\right)\right]\right\}^{-1/2}.
\end{eqnarray}
By integrating over all angular and spectral dependences in either of the formulas~\eqref{eq:angspecdist}, \eqref{eq:angdist} or~\eqref{eq:specdist}, one obtains the total power received from an electron in helical motion as
\begin{equation}\label{eq:totalpower}
    P = \frac{e^4 B^2 \gamma^2\gamma_\parallel^2}{6\pi\epsilon_0 m^2c}\beta_\perp^2 (1-\beta\cos\thetap\cos\mu).
\end{equation}

A number of properties of synchrotron radiation are interesting to note from these formulas. It can be shown that the angular width of the emitted radiation scales as
\begin{eqnarray}
    \psi_\mathrm{c} &= \frac{1}{\gamma}\left(\frac{2\lambda}{\lambdac}\right)^{1/3},\qquad &\lambda\gg\lambdac,\\
    \psi_\mathrm{c} &= \frac{1}{\gamma}\sqrt{\frac{2\lambda}{3\lambdac}},\qquad &\lambda\ll\lambdac,
\end{eqnarray}
in the high and low wavelength limits, respectively. On average, the angular width of the emitted radiation scales as $\psi_\mathrm{c}\sim\gamma^{-1}$, as is realized from the shaping factor $(1-\beta\cos\psi)\approx\frac{1}{2}(\gamma^{-2}+\psi^2)$ of~\Eq{eq:angdist}. For a highly relativistic electron ($\gamma\gg 1$), essentially all synchrotron radiation will be emitted in a very small angular interval around $\psi = 0$, and thus almost entirely in the particle's forward direction. In the guiding-center picture, this corresponds to a cone with opening angle $\thetap$ and lateral width $\sim\gamma^{-1}$ being emitted around the guiding-center. This strong forward-beaming of synchrotron radiation, in combination with the fact that all runaways move in the same direction around the tokamak, also has the consequence that synchrotron radiation from runaway electrons is only detected on one side of the tokamak. This asymmetry in the detected synchrotron radiation between both sides of the tokamak is often used to confirm the presence of runaway electrons in an experiment.

Another interesting property of synchrotron radiation is that its spectrum consists of many closely spaced spectral lines near the peak wavelength $\lambdac$ and so is well described as a continuum by~\Eq{eq:specdist}. Analyzing its asymptotic expressions reveals that the emitted power per wavelength decreases exponentially to short wavelengths ($\lambda\ll\lambdac$) and at a slower $\lambda^{-2/3}$ rate to longer wavelengths.

Most of the synchrotron radiation will be emitted at wavelengths near $\lambda = \lambdac\sim 1 / (\gamma B\sin\thetap)$. We thus see that the spectrum peak can be pushed towards shorter wavelengths in three different ways, namely by either increasing the particle energy, pitch angle or the magnetic field strength. During the transit of an electron in a tokamak, the particle energy will remain constant\footnote{On the transit timescale, in which we are interested, both collisions and radiation losses are negligible.}, but both the magnetic field and pitch-angle will increase when the particle approaches the inboard side of the device. As such, the peak of the particle's synchrotron spectrum will vary during the course of an orbit, which will be demonstrated in Section~\ref{sec:comparisons_energy}, and depending on where the spectrum peak $\lambdac$ is located relative to the visible spectral range of the camera, significantly different contributions of synchrotron radiation from the inboard and outboard sides of the device may be obtained.

\subsection{Quantitative descriptions of synchrotron spots}\label{sec:cone_model}
While the theory presented so far can be used to study many aspects of synchrotron radiation from runaway electrons, this paper will focus mainly on synchrotron images and to some extent also synchrotron spectra. The spots of synchrotron radiation observed in synchrotron images can take on many different shapes, and as was shown in Ref.~\cite{Zhou2014} (and will be further demonstrated in Section~\ref{sec:comparisons}) the observed spot shape is strongly dependent on the placement of the synchrotron camera. In the literature, elliptical synchrotron spots appear to be the most common~\cite{Finken1990,Zhou2013,Yu2013,Wongrach2014,Tong2016}; however, as is exemplified in Sections~\ref{sec:comparisons} and~\ref{sec:Alcator_comparison}, synchrotron spots in Alcator C-Mod tend to take on more crescent-like shapes. Our simulations suggest this is a combined effect due to camera placement and the fairly small pitch angles of the runaways, as will be discussed in Section~\ref{sec:comparison_emission_models}.

To gain a better understanding for how different synchrotron spot shapes arise, and to derive a computationally more efficient model of synchrotron radiation, we can use the strong beaming of synchrotron radiation into a cone of width $\sim\gamma^{-1}$. We let
\begin{equation}
    \frac{\d^2P}{\d\lambda\d\Omega} = \frac{1}{2\pi} \left\langle\frac{\d P}{\d\lambda}\right\rangle \delta\left( 1-\cos\alpha \right),
\end{equation}
where $\alpha$ is the angle between the particle's velocity vector and the observer direction. In terms of the angle $\mu$ between the guiding-center velocity and observer direction, as well as pitch angle $\thetap$ and gyrophase $\zeta$, we can write $\cos\alpha = \cos\mu\cos\thetap + \sin\mu\sin\thetap\cos\zeta$. Representing the delta function in terms of Legendre polynomials $P_l(x)$, and utilizing the addition theorem for spherical harmonics, this implies
\begin{eqnarray}
     \frac{\d^2P}{\d\lambda\d\Omega} = \frac{1}{2\pi} \frac{\d P}{\d\lambda}&\sum_{l=0}^\infty \frac{2l+1}{2}\Biggr[ P_l(\cos\mu) P_l(\cos\thetap) \nonumber \\
    & + 2\sum_{m=1}^l \frac{(-1)^m(l-m)!}{(l+m)!}P_l^m(\cos\mu)P_l^m(\cos\theta_p) \cos m\zeta \Biggr].
\end{eqnarray}
The gyro-average of this expression eliminates the sum over $m$, resulting in
\begin{equation}\label{eq:conemodel}
    \left\langle \frac{\d^2P}{\d\lambda\d\Omega} \right\rangle = \frac{1}{2\pi} \frac{\d P}{\d\lambda} \delta(\cos\mu-\cos\thetap).
\end{equation}
This motivates the idea that the electron guiding-center emits synchrotron radiation in a cone of opening angle $\thetap$ around its direction of motion and gives a simple expression for the integral over the detector surface in~\eqref{eq:softfull}. This simplified model of synchrotron radiation, which we call the \emph{cone model}, has also been implemented in SOFT, and as will be demonstrated in the next section, shows good agreement with the full models. The model also brings computational benefits, and in typical runs it is commonly faster than the full models by a factor between 2-100\footnote{The large variation in time differences between the models is due to that the cone model allows a much stricter sorting algorithm to be employed, which efficiently excludes particles that cannot be detected. The different models therefore scale differently with resolution parameters.}.

Aside from being a useful computational tool, the cone model provides insight into how to interpret the synchrotron spot observed in images. To zeroth order in the guiding-center approximation, $\cos\mu = \mathbf{V}\cdot\mathbf{r}/Vr = \hat{\mathbf{b}}\cdot\mathbf{r}/r$, where $\mathbf{V} = V\hat{\mathbf{b}}$ is the guiding-center velocity and $\mathbf{r}$ is the vector between the detector and the guiding-center. For a guiding-center to emit towards the detector (which we temporarily consider to be a point), the equation
\begin{equation}
    \left|\hat{\mathbf{b}}\cdot\mathbf{r}/r\right| = \left|\cos\thetap\right|,
\end{equation}
must therefore be satisfied. The solution to this equation is a surface in 3D space, which we will refer to as the \emph{surface-of-visibility}, and its shape strongly depends on the detector position. In Alcator C-Mod, for pitch angles around 0.10-$\SI{0.20}{rad}$, it resembles a saddle or potato chip.

\section{Elements and dependencies of the synchrotron image}
\label{sec:comparisons}
\begin{figure}
    \centering
    \begin{minipage}{0.4\textwidth}
        \begin{overpic}[width=\textwidth]{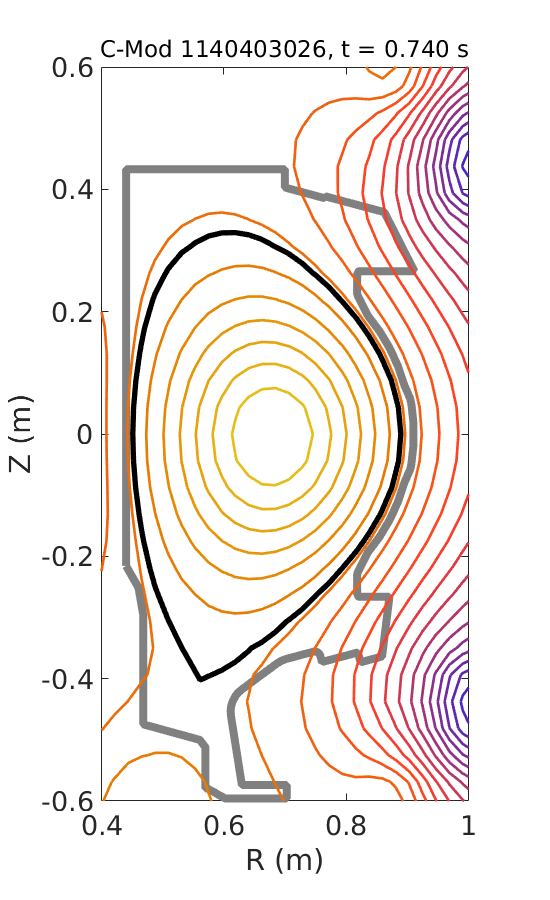}
            \put(13,87){\textbf{(a)}}
        \end{overpic}
    \end{minipage}
    \begin{minipage}{0.59\textwidth}
        \begin{overpic}[width=0.449\textwidth]{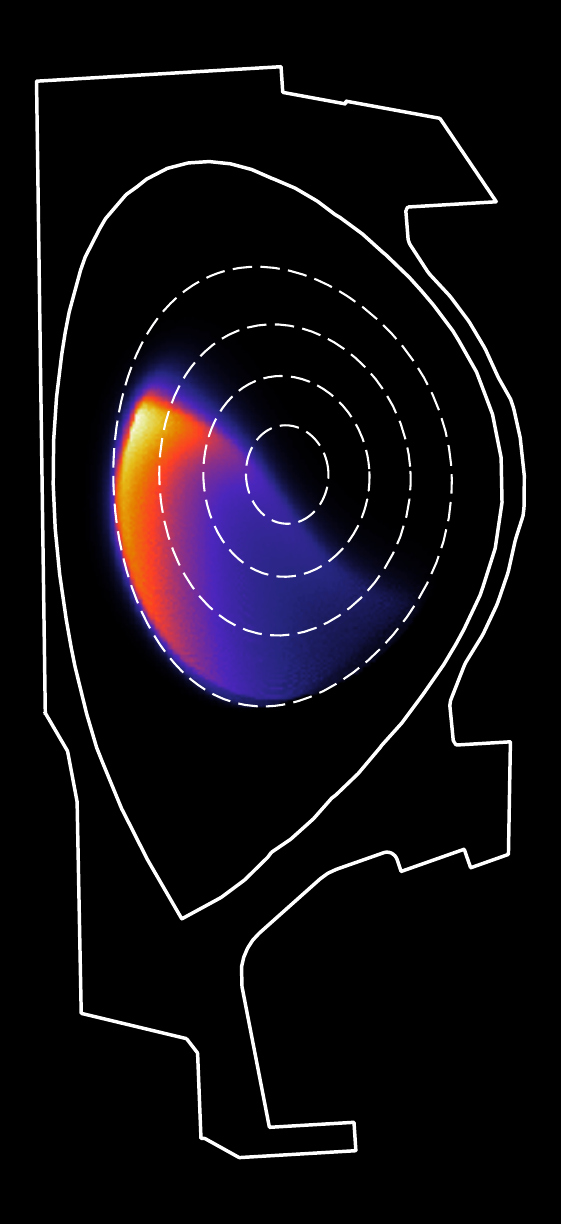}
            \put(37,93){\white{(b)}}
        \end{overpic}
        \begin{overpic}[width=0.49\textwidth]{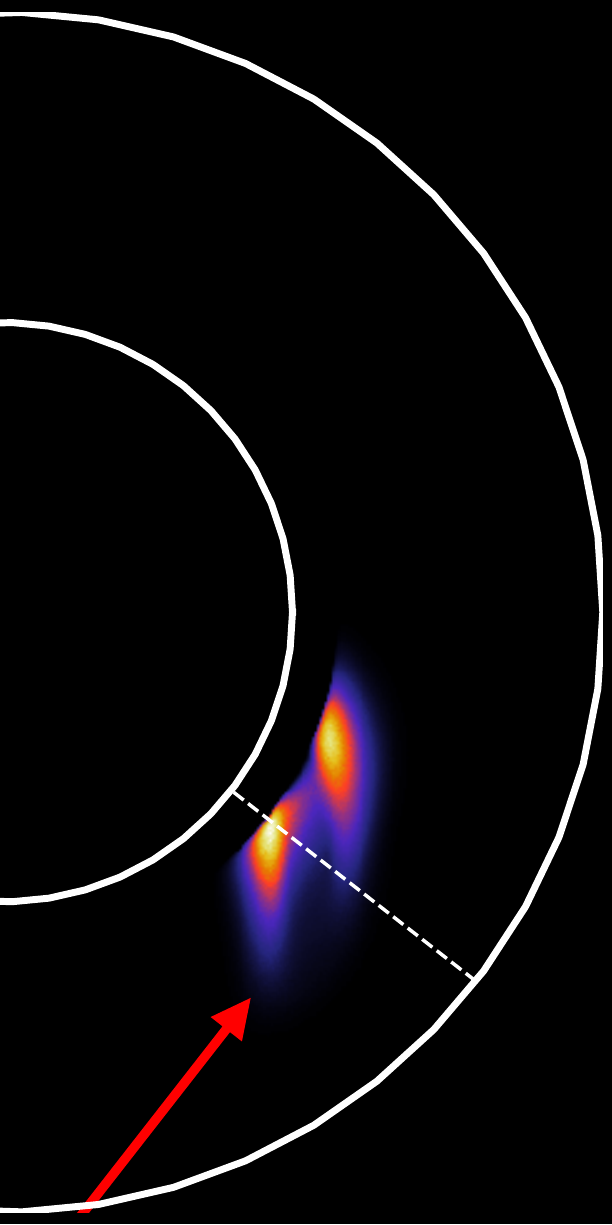}
            \put(40,93){\white{(c)}}
        \end{overpic}
    \end{minipage}
    \caption{(a) Poloidal magnetic flux contours from EFIT \cite{Lao1985} are shown on a poloidal cross-section of the Alcator C-Mod tokamak. The vacuum vessel wall is bolded grey and the plasma boundary bolded black. (b) Example of a synthetic synchrotron image with flux contours and vessel wall superimposed. (c) Top view of the tokamak, showing the spatial origin of the (synthetic) radiation. The white line crossing the synchrotron spot indicates the plane orthogonal to the camera viewing direction (red arrow), which is used as the projection plane in (b).}
    \label{fig:fluxContours}
\end{figure}

Due to the strong dependence on energy, pitch-angle and particle location in both the synthetic detector equation~\eqref{eq:soft} and synchrotron formulas~\eqref{eq:angspecdist}, \eqref{eq:angdist} and \eqref{eq:specdist}, we expect the synchrotron image observed by a camera to similarly depend strongly on these parameters. In this section we will vary a number of key parameters in order to illustrate the effect they have on the synchrotron spot. Specifically, we will use a set of mono-energetic and mono-pitch-angle distributions to study the effects of particle energy, pitch angle, radial distribution and camera vertical position. With the intent of applying the knowledge gained in this section to a more complete and realistic scenario in Section~\ref{sec:Alcator_comparison}, we use Alcator C-Mod parameters from discharge 1140403026 $(t\sim\SI{0.742}{s})$ in these parameter scans. The magnetic topology used is shown in Fig.~\ref{fig:fluxContours}(a) and unless otherwise noted, the synchrotron camera is located at major radius $R\approx\SI{107}{cm}$ and $\sim\SI{20}{cm}$ below the midplane, reflecting its actual position in the device. Fig.~\ref{fig:fluxContours}(b) shows a typical synthetic Alcator C-Mod synchrotron image with flux surfaces and the wall cross-section superimposed, while Fig.~\ref{fig:fluxContours}(c) shows a top view of the tokamak, indicating the spatial extent of the synchrotron emission. Note, that the line-integration is in the vertical direction in the top view, which results in two separate bright features. The white line crossing the synchrotron spot in Fig.~\ref{fig:fluxContours}(c) is the plane orthogonal to the camera viewing direction (indicated by the red arrow), which is also the plane used for projecting the overlays in Fig.~\ref{fig:fluxContours}(b).

Unless otherwise stated, all particles in the following analysis are initiated in a radial interval spanning from the magnetic axis (located at major radius $R = \SI{68}{cm}$ in the magnetic equilibrium used) and $\SI{16}{cm}$ outwards on the outboard side, so that the outermost particle considered is launched at $R = \SI{84}{cm}$. We will also separate the radial dependence from the overall distribution function so that the total distribution function can be written $f(\rho, p_\parallel^{(0)}, p_\perp^{(0)}) = f_r(\rho)f_{\mathbf{p}}(p_\parallel^{(0)},p_\perp^{(0)})$, where $f_r(\rho)$ is the radial profile and $f_{\mathbf{p}}$ is the momentum-space distribution function. Particles are given an initial energy of $E^{(0)} = \SI{30}{MeV}$ and pitch angle $\thetap^{(0)} = \SI{0.15}{rad}$ (i.e.\ $f_{\mathbf{p}}(E^{(0)}, \thetap^{(0)}) = \delta(E^{(0)}-\SI{30}{MeV/c})\delta[\cos\thetap^{(0)}-\cos(0.15)]$), consistent with estimates made for runaways observed in Alcator C-Mod~\cite{Tinguely2015}, and a uniform radial profile is used ($f_r(\rho) = 1$ for $R < \SI{84}{cm}$ and zero otherwise). Only radiation emitted at wavelengths between $\lambda = \SI{500}{nm}$ and $\lambda = \SI{1000}{nm}$ is seen by the detector. The current runs in the counter-clockwise direction (when looking down on the tokamak from above), so that due to the strong forward-beaming of the synchrotron radiation, emission will only be seen on the right side of the tokamak. In most of the images to be presented, the camera will be zoomed in on the synchrotron spot, and the high-field side will be on the left side of the image.

\subsection{Comparison of synchrotron images using different emission models}
\label{sec:comparison_emission_models}

\begin{figure}[h!]
\begin{center}
    \begin{overpic}[width=0.24\textwidth]{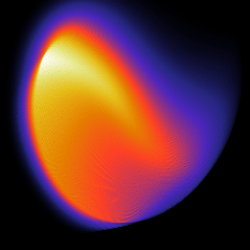}
        \put(80,86){\white{(a)}}
        \put(3,13){\white{Cone model}}
        \put(3,3){\white{w/o spectrum}}
    \end{overpic}
    \begin{overpic}[width=0.24\textwidth]{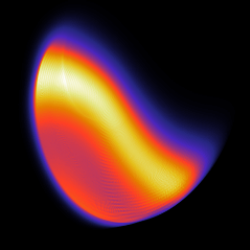}
        \put(80,86){\white{(b)}}
        \put(3,13){\white{Angular}}
        \put(3,3){\white{distribution}}
    \end{overpic}
    \begin{overpic}[width=0.24\textwidth]{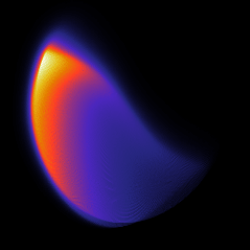}
        \put(80,86){\white{(c)}}
        \put(3,13){\white{Cone model}}
        \put(3,3){\white{w/ spectrum}}
    \end{overpic}
    \begin{overpic}[width=0.24\textwidth]{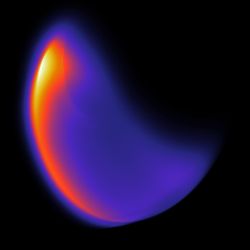}
        \put(80,86){\white{(d)}}
        \put(3,13){\white{Angular \&}}
        \put(3,3){\white{spectral distrib.}}
    \end{overpic}
    \\
    \includegraphics[width=\textwidth]{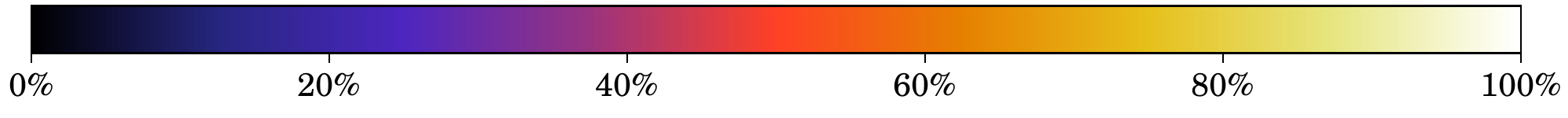}
\end{center}
    \caption{Comparison of synchrotron images using different emission models implemented in \SOFT. The energies and pitch angles of particles are determined by a simulated distribution function, to be described in more detail in Section~\ref{sec:analysis}, and launched on the radial interval $r\in [68, 84]\,\si{cm}$.
    Figure (a) was generated using the cone model by recording the radiation at all wavelengths. Figure (b) takes the angular distribution of radiation Eq.~\eqref{eq:angdist} into account, assumes the radiation to be emitted uniformly across all wavelengths, and records radiation at all wavelengths. Figure (c) shows the result of using the cone model but only registering radiation emitted at wavelengths between 500-$\SI{1000}{nm}$. Figure (d) takes the full angular and spectral distribution~\Eq{eq:angspecdist} into account, and only records radiation received at wavelengths between 500-$\SI{1000}{nm}$. The colors in the figures represent the intensity detected by the camera, normalized  to the most intense pixel of each image.
    }
    \label{fig:models}
\end{figure}
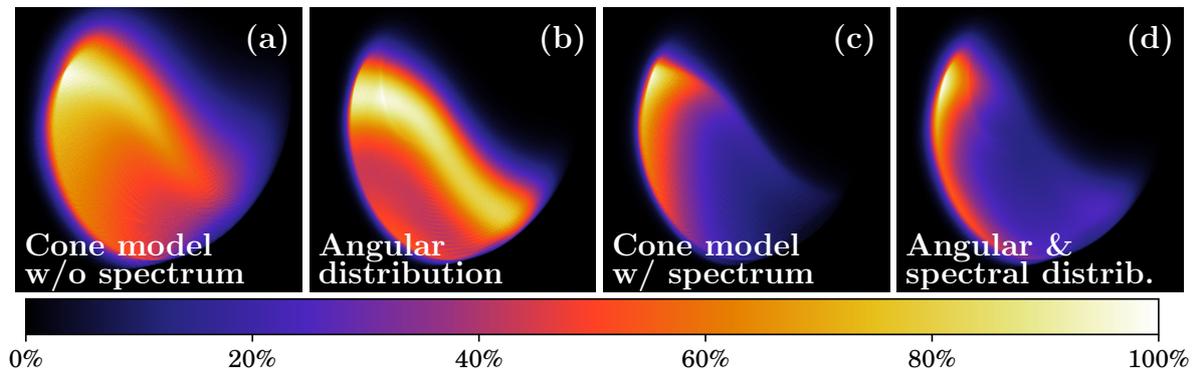

Figure~\ref{fig:models} shows \SOFT~output generated using each of the radiation models discussed in Section~\ref{sec:radiation_models}, simulated with the numerical distribution of runaway electrons to be described in Section~\ref{sec:analysis}. Figures~\ref{fig:models}(a) and~\ref{fig:models}(c) were generated using the approximate cone model Eq.~\eqref{eq:conemodel}, with~\Fig{fig:models}(c) taking the limited spectral range of the camera into account. Figures~\ref{fig:models}(b) and~\ref{fig:models}(d) take the full angular distribution of radiation into account, with the former using just the angular distribution formula~\eqref{eq:angdist} while the latter also respects the camera's limited spectral range with the angular and spectral distribution~\eqref{eq:angspecdist}. As is evident from comparing the figures, taking the limited spectral range into account is important not just for obtaining the correct intensity distribution over the synchrotron spot, but also for obtaining the correct overall spot shape. Comparing Figs.~\ref{fig:models}(c)-(d) also reveals that the cone model is accurate in describing the emitted synchrotron radiation, and it will hence be used in the rest of this paper due to its computational efficiency.

\subsection{Contributions from individual radii}\label{sec:comparisons_radii}
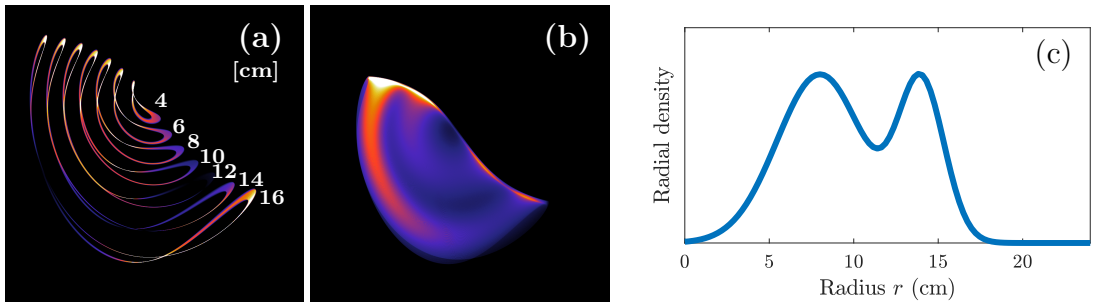
\begin{figure}[h]
    \begin{center}
        \begin{overpic}[width=0.25\textwidth]{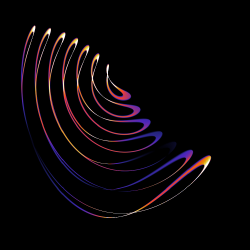}
            \put(78,86){\white{(a)}}
            \put(76,76){\scriptsize{\white{[cm]}}}
            \put(50,66){\scriptsize\white{4}}
            \put(56,57){\scriptsize\white{6}}
            \put(61,52){\scriptsize\white{8}}
            \put(65,47){\scriptsize\white{10}}
            \put(69,42){\scriptsize\white{12}}
            \put(78,39){\scriptsize\white{14}}
            \put(85,33){\scriptsize\white{16}}
        \end{overpic}
        \begin{overpic}[width=0.25\textwidth]{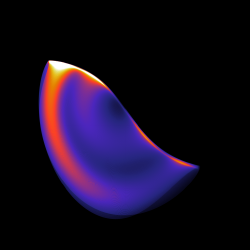}
            \put(78,86){\white{(b)}}
        \end{overpic}
        \begin{overpic}[height=0.25\textwidth]{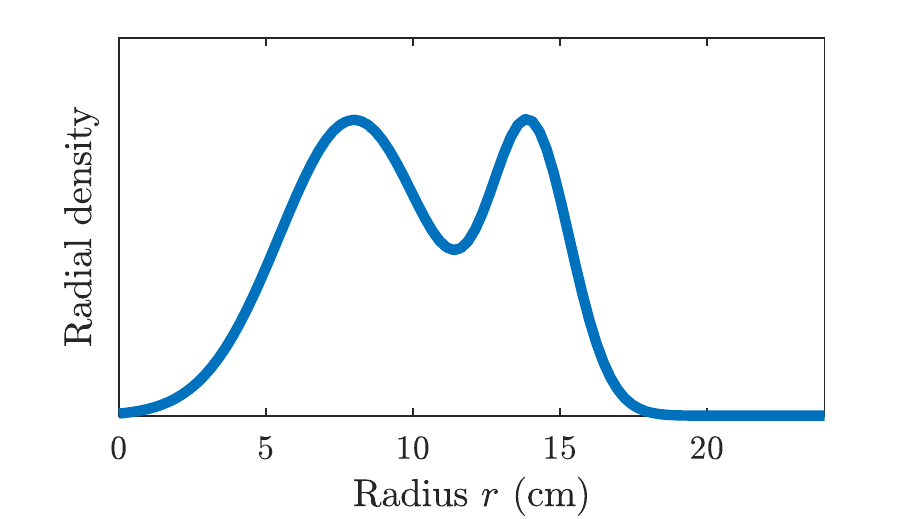}
            \put(80,46){(c)}
        \end{overpic}
    \end{center}
    \caption{Radiation emission from particles launched at different radii. Figure (a) shows the result of launching particles from a set of eight uniformly spaced points from the magnetic axis ($R=\SI{68}{cm}$) and $\SI{16}{cm}$ outwards ($R = \SI{84}{cm}$). Contributions from the particles closest to the magnetic axis are not visible to the camera due to its vertical displacement. The numbers denote the distance, in cm, from the magnetic axis at which the particles were launched. In figure (b) the radial profile shown in figure (c), consisting of two Gaussian functions, was used to illustrate the effect of having a non-uniform spatial distribution of particles.}
    \label{fig:radial_dist}
\end{figure}
As was discussed in Section~\ref{sec:synthdiagtheory}, the radial parameter in \SOFT\ describes the major radius on the midplane at which the particle's guiding-center initiates its orbit, and because of the toroidal symmetry of the system and Liouville's theorem, it is the only spatial parameter of the distribution function. To reveal how contributions from a particle at a given radius appear in a synchrotron image, particles were initialized every $\SI{2}{cm}$ in the outer midplane, from the magnetic axis and $\SI{16}{cm}$ outwards, and the resulting image is shown in Fig.~\ref{fig:radial_dist}(a). Note that these results are specific to the case when the camera is located far below the midplane, although a similar but more symmetric behaviour is found when the camera is located in the midplane.

The contribution to the image from a particle launched at a single radius can be described as a closed ribbon, extending along a parabola, from the upper left of the image to the middle-right. Particles initialized at larger radii contribute longer ribbons of radiation, each appearing further to the left (high-field side) in the image. The distinct radius plot \Fig{fig:radial_dist}(a) reveals how a radial distribution would affect the synchrotron spot, since each ribbon would be weighted differently, and in \Fig{fig:radial_dist}(b) a particular radial profile consisting of two Gaussian curves has been used, as shown in Fig.~\ref{fig:radial_dist}(c). The use of a radial profile consisting of two Gaussian functions illustrates how a peaked radial profile affects the synchrotron spot, which is of interest to the analysis of the experimental image in Section~\ref{sec:analysis}.

Another interesting effect seen from Fig.~\ref{fig:radial_dist} is that despite particles being initialized on the magnetic axis and at $r = \SI{2}{cm}$, contributions to the image from those radii do not appear in the image. This is a combined effect of the camera being located $\sim\SI{20}{cm}$ below the midplane and the pitch angles of the particles being too small, so that particles moving close to the magnetic axis (which move essentially along it) never emit radiation towards the camera. While this effect is not necessarily significant in a camera image, it may strongly impact the observed synchrotron spectrum as, effectively, the \emph{observed} runaway distribution function gets biased in favour of particles with large pitch angles.

\subsection{Camera position}
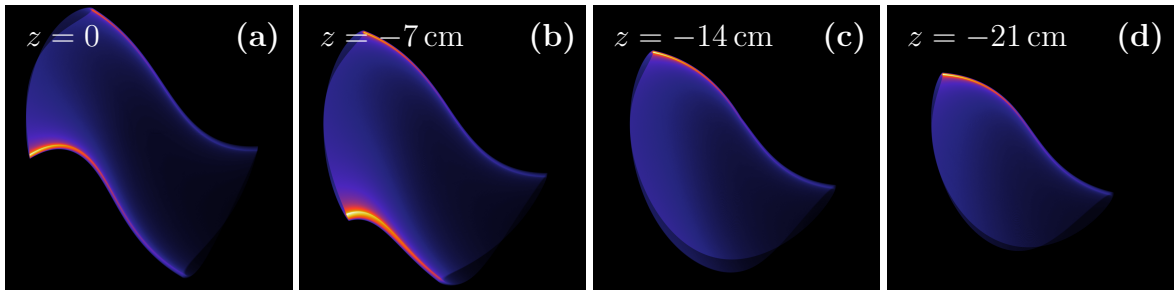
\begin{figure}[h]
    \begin{center}
        \begin{overpic}[width=0.24\textwidth]{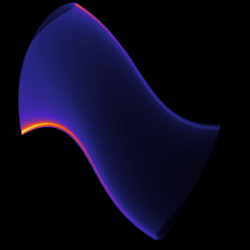}
            \put(7,86){\white{$z = 0$}}
            \put(80,86){\white{(a)}}
        \end{overpic}
        \begin{overpic}[width=0.24\textwidth]{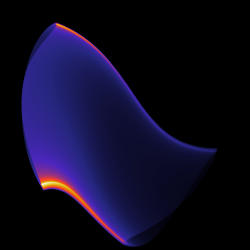}
            \put(7,86){\white{$z = \SI{-7}{cm}$}}
            \put(80,86){\white{(b)}}
        \end{overpic}
        \begin{overpic}[width=0.24\textwidth]{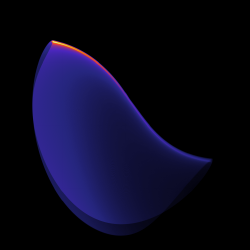}
            \put(7,86){\white{$z = \SI{-14}{cm}$}}
            \put(80,86){\white{(c)}}
        \end{overpic}
        \begin{overpic}[width=0.24\textwidth]{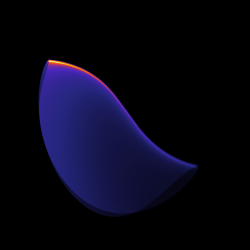}
            \put(7,86){\white{$z = \SI{-21}{cm}$}}
            \put(80,86){\white{(d)}}
        \end{overpic}
    \end{center}
    \caption{Comparison of synchrotron images using different vertical position of the camera. When the camera is placed in the midplane ($z = 0$) the image becomes more symmetric compared to placing it $z = -\SI{21}{cm}$ below the midplane, as is done in Alcator C-Mod.}
    \label{fig:midplane_spots}
\end{figure}
Because of the pitch angle's limiting effect on the extent of the synchrotron spot, the vertical displacement of the camera turns out to greatly impact the synchrotron image. In the images presented so far the camera was located $\sim\SI{20}{cm}$ below the midplane, as in the Alcator C-Mod tokamak, and the resulting synchrotron radiation spot resembles a saddle. When we move the camera to the midplane we obtain the synchrotron radiation spot shown in \Fig{fig:midplane_spots}(a), which has a more symmetric shape. What appeared to be a saddle when viewed from below the midplane now becomes a twisted cylinder.

The bright edges appearing on most single-particle synchrotron spots are results of the three-dimensional extent of the synchrotron emission. As discussed in Sec.~\ref{sec:cone_model}, particles can only emit towards the detector in certain points of space, and we refer to this set of points as the \emph{surface-of-visibility}. When the detector has a finite size, as is the case in our simulations, the surface turns into a thin volume with the same overall shape. Along certain edges of the volume, the volume curves, and a line-of-sight from the detector may for a small section of its path extend tangentially through the volume. These parts of the spot tend to be significantly brighter than any other parts.

\subsection{Effect of pitch angle}\label{sec:comparisons_pitch}
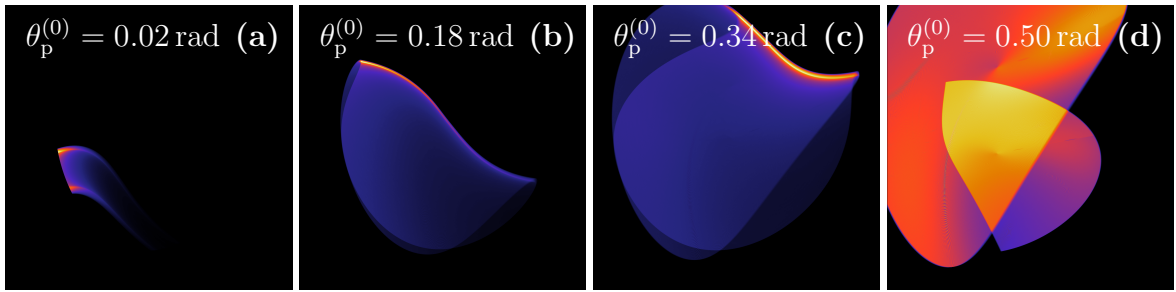
\begin{figure}[ht]
    \begin{center}
        \begin{overpic}[width=0.24\textwidth]{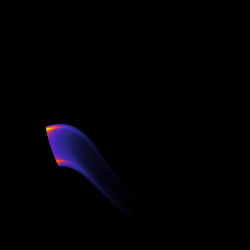}
            \put(7,86){\white{$\thetap^{(0)} = \SI{0.02}{rad}$}}
            \put(80,86){\white{(a)}}
        \end{overpic}
        \begin{overpic}[width=0.24\textwidth]{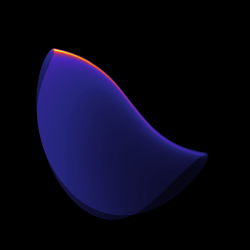}
            \put(7,86){\white{$\thetap^{(0)} = \SI{0.18}{rad}$}}
            \put(80,86){\white{(b)}}
        \end{overpic}
        \begin{overpic}[width=0.24\textwidth]{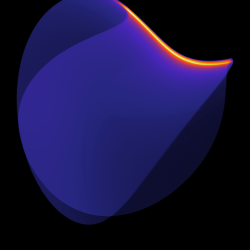}
            \put(7,86){\white{$\thetap^{(0)} = \SI{0.34}{rad}$}}
            \put(80,86){\white{(c)}}
        \end{overpic}
        \begin{overpic}[width=0.24\textwidth]{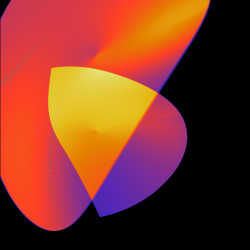}
            \put(7,86){\white{$\thetap^{(0)} = \SI{0.50}{rad}$}}
            \put(80,86){\white{(d)}}
        \end{overpic}
    \end{center}
    \caption{Comparison of synchrotron images using particles with the same energy $E^{(0)} = \SI{30}{MeV}$ but different pitch angles. The pitch angle of the particles determines the ``width'' of the synchrotron radiation spot. At small pitch angles, a saddle-like surface is observed, while at large pitch angles it separates into two distinct oval shapes. Note that all images are normalized to their maximum brightness. The brightest features of (a)-(c) are due to ``line-integration'' effects of the surface-of-visibility, whereas (d) lacks this effect since it consists of two separated surfaces.}
    \label{fig:param_pitch}
\end{figure}
The pitch angle of the particle mainly determines the width of the synchrotron spot, as shown in \Fig{fig:param_pitch} and discussed in for example Refs.~\cite{Finken1990,Jaspers1995,Pankratov1996}. The spot shape can be understood by considering the cone model of synchrotron emission, which for small $\thetap$ suggests that the guiding-center should emit radiation in a very narrow cone in its direction of motion. The particle is only seen when a part of the cone surface intersects the camera, which is less likely to occur when the pitch-angle is small.

The shape of the synchrotron spot at small pitch angles is a saddle-like surface, as seen in Figs.~\ref{fig:param_pitch}(a)-\ref{fig:param_pitch}(c). At large pitch angles, the saddle-like surface opens up along its upper edge and separates into two distinct oval surfaces that start to move away slowly from each other as the pitch angle further increases, as shown in Fig.~\ref{fig:param_pitch}(d). The apparent brightness of Fig.~\ref{fig:param_pitch}(d) is due to that all images are normalized to their brightest point. Since the surface-of-visibility has split into two distinct surfaces in Fig.~\ref{fig:param_pitch}(d), it lacks a very bright line-integrated contribution which the surface-of-visibilities of Figs.~\ref{fig:param_pitch}(a)-(c) have along their upper edges.

\subsection{Effect of energy}\label{sec:comparisons_energy}
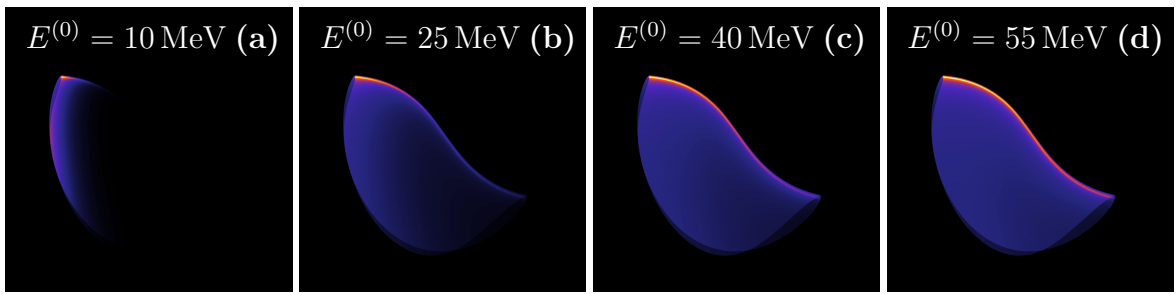
\begin{figure}[h]
    \begin{center}
        \begin{overpic}[width=0.24\textwidth]{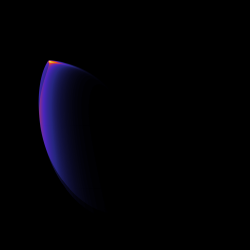}
            \put(7,86){\white{$E^{(0)} = \SI{10}{MeV}$}}
            \put(80,86){\white{(a)}}
        \end{overpic}
        \begin{overpic}[width=0.24\textwidth]{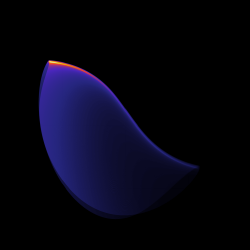}
            \put(7,86){\white{$E^{(0)} = \SI{25}{MeV}$}}
            \put(80,86){\white{(b)}}
        \end{overpic}
        \begin{overpic}[width=0.24\textwidth]{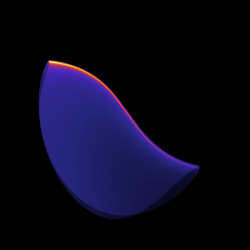}
            \put(7,86){\white{$E^{(0)} = \SI{40}{MeV}$}}
            \put(80,86){\white{(c)}}
        \end{overpic}
        \begin{overpic}[width=0.24\textwidth]{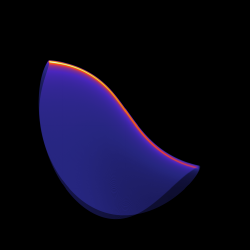}
            \put(7,86){\white{$E^{(0)} = \SI{55}{MeV}$}}
            \put(80,86){\white{(d)}}
        \end{overpic}
    \end{center}
    \caption{Comparison  of  synchrotron  images  using particles with different energies and the same initial pitch-angle $\thetap^{(0)} = \SI{0.13}{rad}$. A particle's energy determines both the focus of intensity in the image, as well as the maximum width of the synchrotron spot.}
    \label{fig:param_energy}
\end{figure}

\begin{figure}
    \centering
    \includegraphics[width=0.5\textwidth]{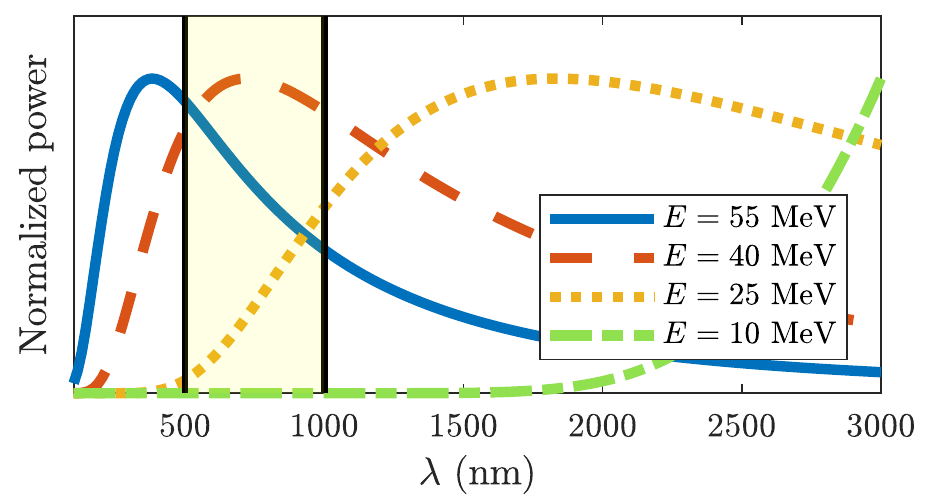}
    \caption{Synchrotron spectra, given by~\Eq{eq:specdist}, for various particle energies, normalized to the maximum value of each spectrum. At fixed magnetic field $B=\SI{5.25}{T}$ and pitch angle $\thetap = \SI{0.13}{rad}$, the peak of the synchrotron spectrum is pushed to shorter wavelengths as the particle energy is increased. The spectral range visible to the camera is marked by two vertical black lines. The sharp fall-off of the spectrum in the visible range at low energy is the reason for the behaviour of the synchrotron spot observed in Fig.~\ref{fig:param_energy}.}
    \label{fig:energy_spectra}
\end{figure}

The main effect of the particle energy is related to the limited spectral range of the camera, since the energy does not appear explicitly in~\eqref{eq:conemodel} which determines the spot shape. In Fig.~\ref{fig:param_energy}(a), synchrotron radiation is only seen in the left part of the image, corresponding to the high-field side of the tokamak, and when the particle energy is increased more radiation starts to appear also from the right parts of the images. The cause of this is illustrated in Fig.~\ref{fig:energy_spectra}, which shows the dependence of the synchrotron spectrum from a single particle located on the magnetic axis, on the energy of the particle. The spectral range of the camera considered here, with wavelengths between $\lambda\in[500, 1000]\,\si{nm}$, is marked in the plot with two vertical lines. At the lower energy, $E^{(0)} = \SI{10}{MeV}$, the spectrum peak lies at longer wavelengths than those observed by the camera, and the synchrotron emission depends exponentially on the magnetic field in this region, as shown in Fig.~\ref{fig:energy_spectra}. As the energy is increased, the spectrum peak is pushed towards shorter wavelengths, and thus also closer to the spectral range of the camera.

At the higher energies, particularly $E^{(0)}=\SI{55}{MeV}$, the spectrum peak is located at wavelengths much shorter than those detected by the camera so that radiation is emitted in the opposite limit, $\lambda\gg\lambdac$. In this limit the amount of radiation emitted in the camera's spectral range scales roughly as $(\lambdac/\lambda)^{2/3}\sim B^{-1/3}$ (also taking the dependence of the particle's pitch angle on $B$ into account), and no significant variation in the intensity over the synchrotron radiation spot is observed between the high- and low-field sides.

\section{Application: Analysis of a synchrotron image in Alcator C-Mod}
\label{sec:Alcator_comparison}

\subsection{Experimental observations}
\label{sec:experiment}
\begin{figure}
    \centering
    \includegraphics[width=0.6\textwidth]{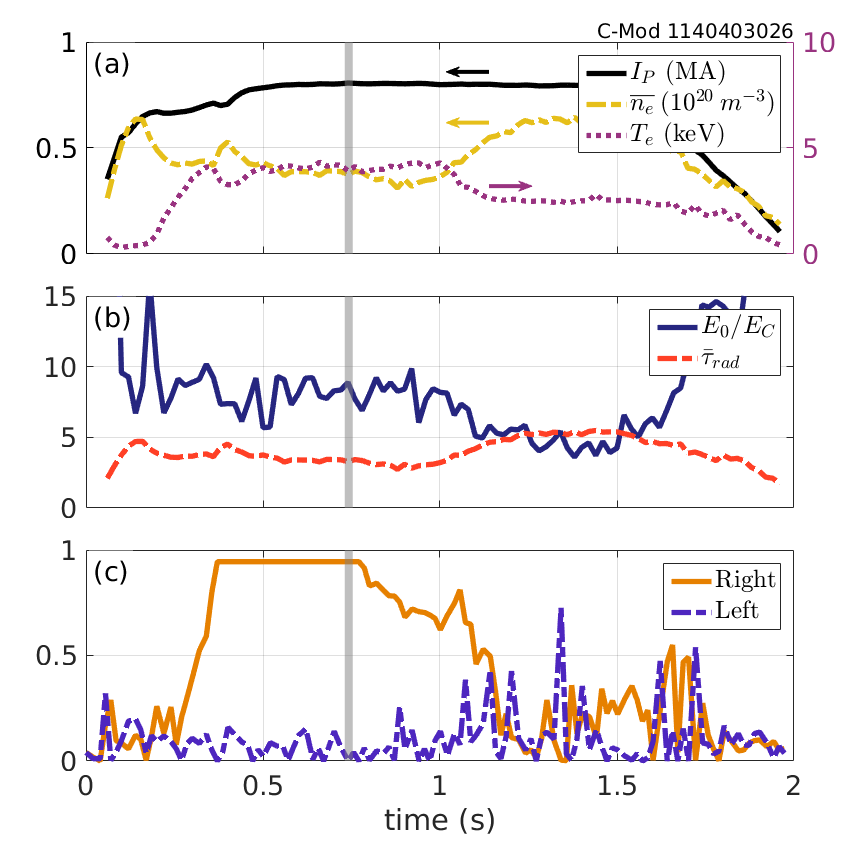}
    \caption{Time traces are plotted for C-Mod shot 1140403026 with time of interest $t\sim \SI{0.742}{s}$ indicated by the grey vertical line: (a) Plasma current $I_P$ (solid), line-averaged electron density $\overline{n_e}$ (dashed), and central electron temperature $T_e$ (dotted); (b) Ratios of the electric field on axis $E_0$ to the critical electric field $E_c$ (solid) and radiation timescale to collision time $\hat{\tau}_{\mathrm{rad}} = \tau_{\mathrm{rad}}/\tau_c$ (dashed); and (c) the intensities as measured by a camera on the Right (solid) and Left (dashed) sides of Figure \ref{fig:camera1140403026}. The toroidal magnetic field on axis was $\SI{5.4}{T}$.}
    
    \label{fig:plasmaParameters}
\end{figure}

\begin{figure}
    \centering
    \includegraphics[width=0.6\textwidth]{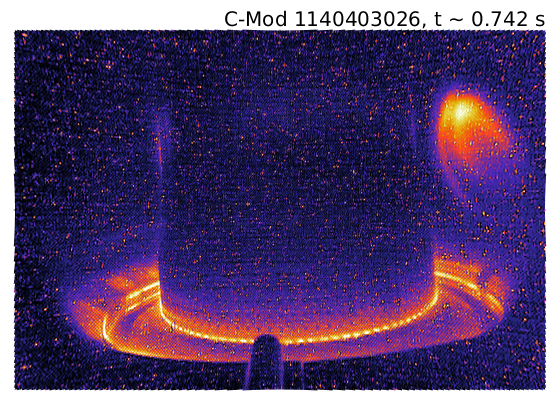}
    \caption{A camera inside C-Mod captures the spatial pattern of visible synchrotron emission during discharge 1140403026 at time t $\sim \SI{0.742}{s}$. Note that this image has been corrected for camera lens distortion, and a perceptually-uniform colormap has been applied to highlight details while conserving camera intensity. The plasma parameters and intensities measured by the camera are shown in Fig. \ref{fig:plasmaParameters}.}
    \label{fig:camera1140403026}
\end{figure}

The Alcator C-Mod tokamak is a high field, compact fusion experiment with major and minor radii of $\SI{68}{cm}$ and $\SI{22}{cm}$, respectively. Relativistic runaway electrons can be generated in low density C-Mod plasma discharges; however, they are not observed after disruptions, likely due to the fast break up of magnetic flux surfaces \cite{Izzo2011}. Instead, electrons will run away to relativistic speeds when the line-averaged electron density $\overline{n}_e < 0.5\times10^{20}\,$m$^{-3}$, whereas normal operating densities are $\overline{n}_e \sim 10^{20}\,$m$^{-3}$. Plasma parameters for a runaway-producing C-Mod discharge are shown in Figure \ref{fig:plasmaParameters}. In this particular experiment, the density was low during the first half of the flattop current ($\sim$~0.5--$\SI{1}{s}$); thus, the electric force driving plasma current was sufficient to overcome the collisional friction acting on a small fraction of electrons and accelerate these particles to relativistic energies. Figure \ref{fig:plasmaParameters}(b) shows that the electric field at the magnetic axis, $E_0$, is approximately 5--10 times greater than the critical electric field required to generate runaway electrons \cite{ConnorHastie}, $E_c = n_e e^3 \ln\Lambda/4 \pi \epsilon_0^2 m_e c^2$, which is consistent with empirical evidence \cite{Granetz2014} and theoretical predictions \cite{Stahl2015}. Note also that due to C-Mod's high magnetic field -- 5.4 T for this discharge -- synchrotron radiation is an important power loss mechanism for runaways as indicated by the low values of $\hat{\tau}_{\mathrm{rad}}$, the ratio between radiation and collisional timescales, in Fig. \ref{fig:plasmaParameters}(b). As runaways are generated, a wide-view camera inside C-Mod measures increasing intensity of visible synchrotron emission from the counter-current direction, as seen in Figs. \ref{fig:plasmaParameters}(c) and \ref{fig:camera1140403026}. At $t\sim\SI{1}{s}$, the density is ramped up, leading to a decrease in $E_0/E_C$ and reduction of runaway electrons.
    
A visible-light camera with wide field of view captures images of each C-Mod plasma discharge at approximately 60 frames per second. Because C-Mod operates at high magnetic fields (2--8 T), the peak emission of synchrotron spectra is shifted toward the visible wavelength range. Therefore, these images can provide important measurements of the spatial distribution and evolution of runaway electrons. The analysis is complicated by the interplay of the camera view and magnetic field geometries, as runaways emit synchrotron radiation primarily along their direction of motion. This camera is located at a major radius of $\SI{106.9}{cm}$ and vertical position of $\SI{-22.655}{cm}$ (where the midplane is $z = 0$). The viewing angle is approximately $\SI{3.4}{degrees}$ to the left of the axis of symmetry and $\SI{1.6}{degrees}$ upward from the horizontal. The hemispherical lens of the camera causes barrel distortion of the image, where straight lines in real space appear bent outward in the image. However, in-vessel calibrations were performed to correct for distortion and reproduce a rectilinear image, which can be compared to synthetic diagnostics.

The synchrotron emission patterns observed by the camera can display many interesting features. For example, in Figure \ref{fig:camera1140403026}, two tilted parabolic structures appear to overlap each other, with the brightest emission at higher vertical position and smaller major radius.  The magnetic flux surfaces on which runaway electrons form certainly impact the spatial distribution of runaway electrons and their observed radiation. Poloidal magnetic field coils and toroidal flux loops at many locations around the vacuum vessel are used in C-Mod to reconstruct the magnetic flux geometry inside the vessel and plasma. Contours of flux overlay a poloidal cross-section of the vacuum vessel in Figure \ref{fig:fluxContours}(a), where the plasma boundary is drawn in bolded black.

%

\subsection{Full distribution simulations with SOFT}
\label{sec:analysis}
\begin{figure}
    \centering
    \begin{overpic}[width=0.49\textwidth]{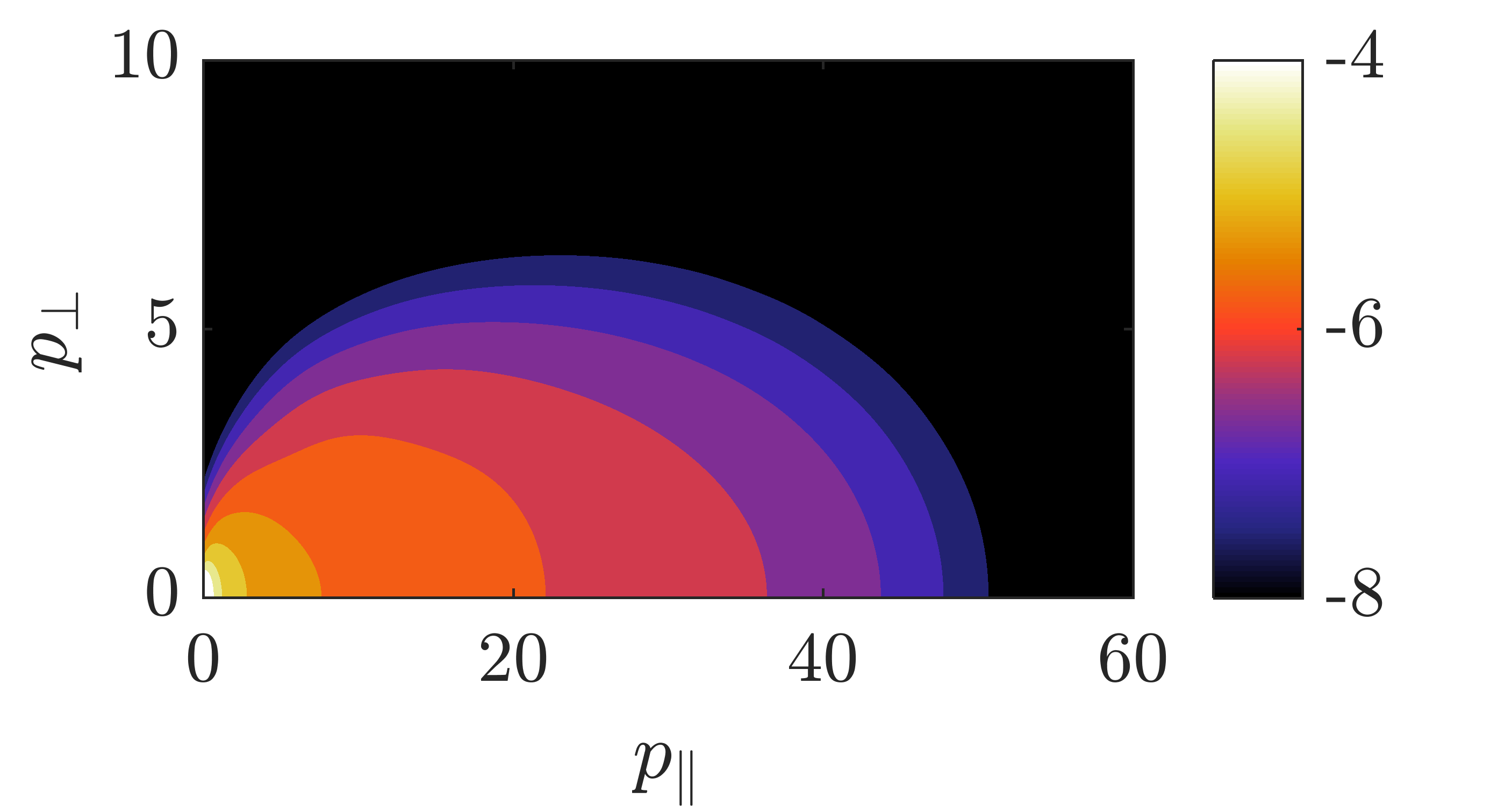}
        \put(16,42){\white{(a)}}
        \put(27,42){\white{$\log_{10} (f / f(p = 0))$}}
    \end{overpic}
    \begin{overpic}[width=0.49\textwidth]{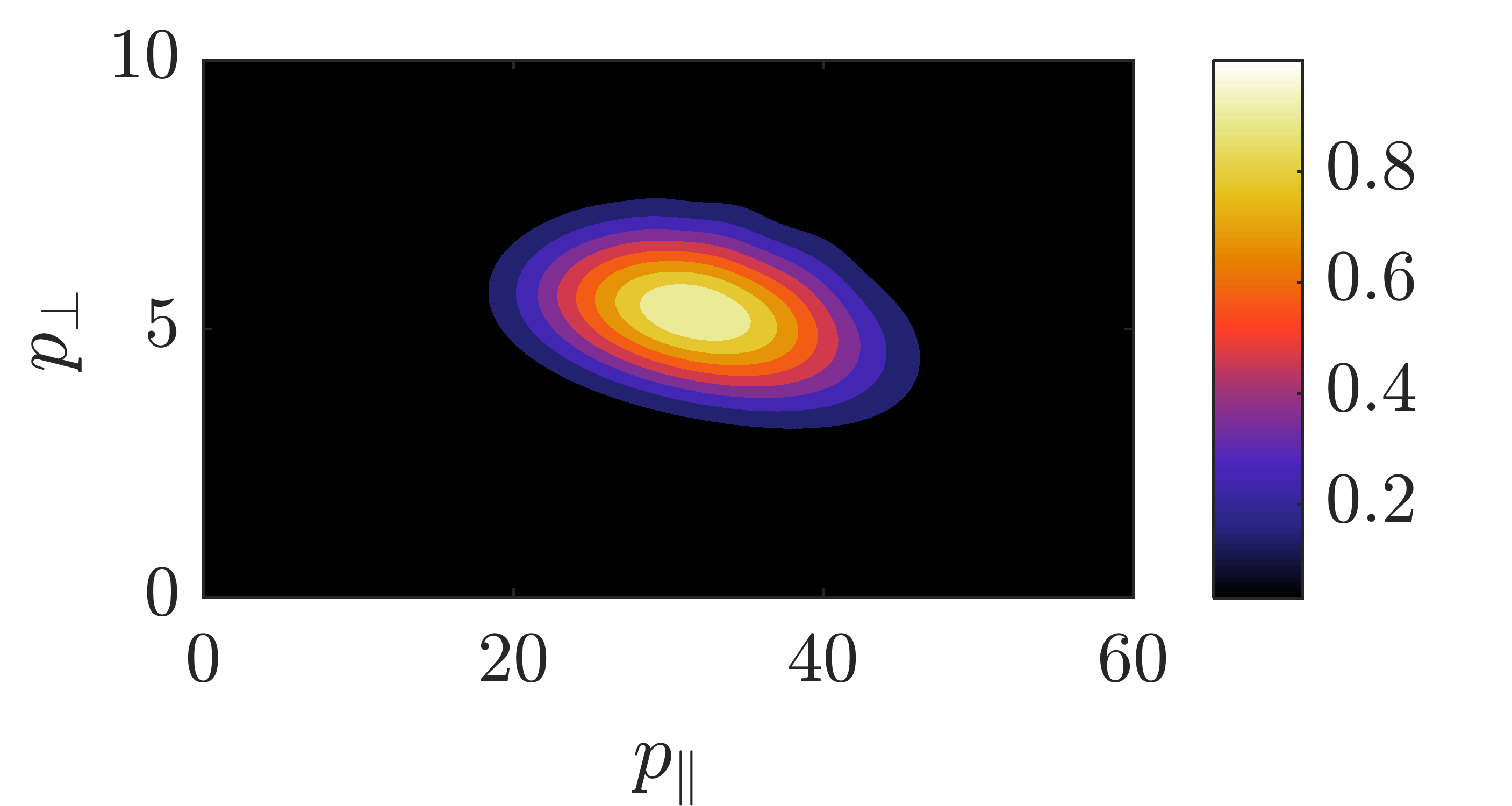}
        \put(16,42){\white{(b)}}
        \put(30,42){\white{$\hat{I}f / \mathrm{max}(\hat{I}f)$}}
    \end{overpic}
    \caption{The momentum-space distribution function used in modelling Fig.~\ref{fig:camera1140403026} is shown in part (a), normalized to its peak value. In part (b), the distribution function has been weighted with the amount of synchrotron radiation $\hat{I}$ emitted in the wavelength interval $[500,1000]\,$\si{\nano\meter} from each point of momentum space, as given by~\Eq{eq:specdist}. The resulting contour plot reveals which parts of momentum space will dominate the spectrum and image for this particular population of runaways.}
    \label{fig:dist}
\end{figure}

In previous sections we considered radially uniform mono-energetic and mono-pitch-angle electron distributions, but the populations of runaway electrons in tokamaks are widely distributed in phase-space and typically far from mono-energetic. To model the image in Fig.~\ref{fig:camera1140403026}, we use a momentum-space distribution of runaway electrons obtained from kinetic simulations. In this work, we use \textsc{CODE}~\cite{Landreman2014,Stahl2016} to solve the spatially homogeneous kinetic equation for electrons in 2D momentum space, including electric-field acceleration, collisions, synchrotron-radiation reaction losses and the knock-on source term given by Chiu et al.~\cite{Chiu1998}. Figure~\ref{fig:dist} shows the runaway electron distribution function for the parameters at the time-slice indicated by the grey line in Fig.~\ref{fig:plasmaParameters}, $t\sim\SI{0.742}{s}$. Note that only parameters on the magnetic axis have been used in generating the distribution function, and therefore the output is only an estimate of the distribution function. To more accurately calculate the distribution function, radial profiles of parameters such as temperature, electron density, electric field, etc.\ would be needed.

As described in Section~\ref{sec:synthdiagtheory}, to utilize the physics of the system as much as possible, the distribution function in SOFT is specified in the outer midplane along the line $\tau = \phi = 0$. From this line, the guiding-center equations of motion are then used to evolve the distribution function in the poloidal plane, and particles are uniformly distributed in the toroidal direction. The integration regions for the variables $\tau$ (orbit time) and $\phi$ are constrained by the physics of the system to be $\tau\in [0, \tau_\mathrm{pol}(\rho, p_\parallel, p_\perp)]$ and $\phi\in [0, 2\pi]$, where $\tau_\mathrm{pol}(\rho, p_\parallel, p_\perp)$ denotes the time it takes for a particle beginning its orbit at major radius $\rho$, with initial momentum $p_\parallel$ and $p_\perp$ along and perpendicular to the magnetic field respectively, to return to its starting point in the poloidal plane. The only integration regions in phase-space that must be specified are therefore the runaway beam size (maximum radius $\rho$), and the 2D momentum-space region from which to sample particles (for $p_\parallel^{(0)}$ and $p_\perp^{(0)}$). We choose a runaway beam size of $\SI{16}{cm}$, placing particles in 100 points between $\rho = 68\text{-}\SI{84}{cm}$; a total of 300 points for $p_\parallel^{(0)}\in[10,25]\,\si{MeV/c}$ and 100 points for $p_\perp^{(0)}\in[1.5, 4]\,\si{MeV/c}$, the latter two based on the region in Fig.~\ref{fig:dist}(b) from which we expect the most significant contributions to the image. The detector is modeled as a square of side $\SI{6}{mm}$ and is assumed to only see radiation emitted at wavelengths between $\lambda\in[500,1000]\,\si{nm}$, with a uniform response function.

\begin{figure}
    \centering
    \begin{overpic}[height=0.25\textwidth]{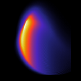}
        \put(10,86){\white{(a)}}
    \end{overpic}
    \begin{overpic}[height=0.25\textwidth]{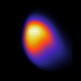}
        \put(10,86){\white{(b)}}
    \end{overpic}
    \begin{overpic}[height=0.25\textwidth]{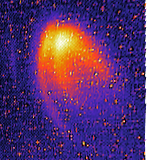}
        \put(5,87){\white{(c)}}
    \end{overpic}
    \caption{(a) Synchrotron image obtained with SOFT using the distribution of runaway electrons shown in Fig.~\ref{fig:dist} and a uniform radial profile. (b) The same momentum-space distribution as in (a), but with a linearly decaying radial profile. (c) Zoomed view of the camera image in Fig.~\ref{fig:camera1140403026}. }
    \label{fig:realistic}
\end{figure}

The synthetic synchrotron images obtained from \SOFT~using the momentum-space distribution function and parameters just described, and with two different radial distributions, are presented in Fig.~\ref{fig:realistic}. In Fig.~\ref{fig:realistic}(a) a uniform radial profile was used, while in Fig.~\ref{fig:realistic}(b) a radial profile decreasing to zero linearly to the plasma edge at $R=\SI{84}{cm}$ was considered. Despite using these rather basic models for the distribution function, the agreement of Figs.~\ref{fig:realistic}(a) and \ref{fig:realistic}(b) with the experimentally obtained camera image Fig.~\ref{fig:realistic}(c) is quite good. The crescent shape is a result of the camera being located $\sim\SI{20}{cm}$ below the midplane, in combination with the dominating pitch angle being $\thetap^{(0)} = \SI{0.15}{rad}$, as indicated by Fig.~\ref{fig:dist}(b). While a linear radial profile significantly increases the similarity between the simulated and experimental image, the double crescent shape of Fig.~\ref{fig:dist}(c) cannot be reproduced without a modulation of the linearly decreasing radial profile. An off-axis peak, combined with a near-linear decrease in the radial profile, could explain the experimentally observed spot-shape. The effect of a spatially varying momentum-space distribution should also be explored.

\begin{figure}
    \centering
    \includegraphics[width=0.65\textwidth]{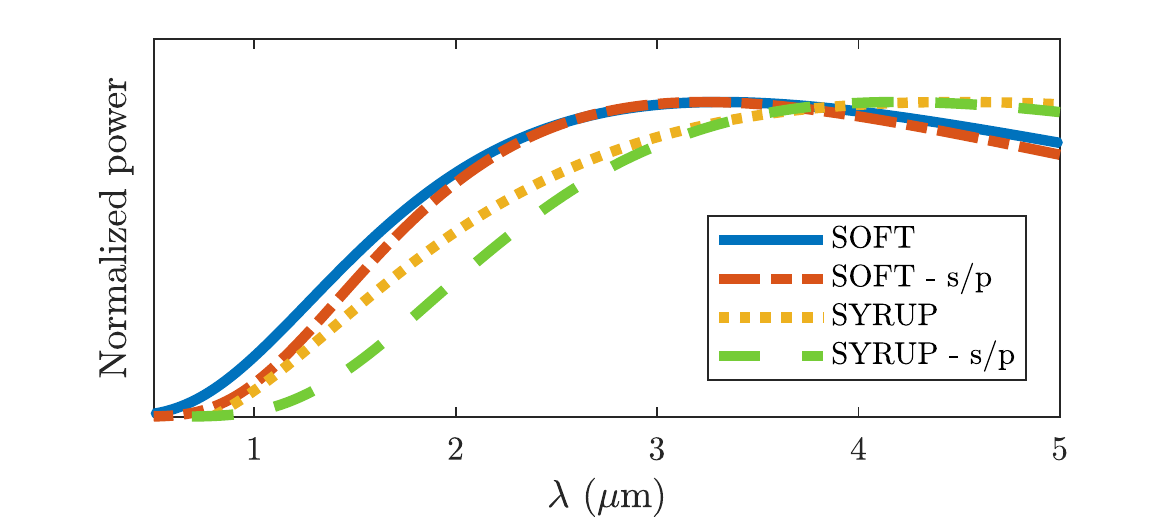}
    \caption{Comparison of synchrotron spectra for the Alcator C-Mod scenario using SOFT and the numerical tool SYRUP (described in Ref.~\cite{Stahl2013}) respectively. The blue solid and yellow dotted lines are the spectra resulting from SOFT and SYRUP respectively when contributions from the entire runaway population is considered. The red dash-dotted and green dashed lines are ``single particle'' (s/p) spectra, generated by locating the source of dominant emission in momentum-space ($E^{(0)}\sim\SI{14}{MeV}$, $\thetap^{(0)}\sim\SI{0.15}{rad}$, see Fig.~\ref{fig:dist}) and computing the corresponding spectra. It should be noted that the SYRUP spectra have been generated with the maximum magnetic field strength experienced by any particle, $B = \SI{7.13}{T}$, which clearly is not sufficient for agreement.
    }
    \label{fig:allspectra}
\end{figure}

Comparing the synchrotron spectra calculated with SOFT and the numerical tool SYRUP~\cite{Stahl2013}, respectively, from the distribution of runaways shown in Fig.~\ref{fig:dist} reveals that geometric effects significantly impact the spectrum. In Fig.~\ref{fig:allspectra}, four spectra relevant to the Alcator C-Mod scenario described in this paper are shown -- two with the full \CODE~distribution function of~\Fig{fig:dist} and two for which the dominating particle of that distribution has been identified and the corresponding spectrum calculated.

While \SOFT~captures the geometry of the device, simulates the detector, and describes the spatial dependence of the radiation, SYRUP computes the synchrotron spectrum from a given momentum-space distribution function. Since the distribution is initialized in the outer midplane in \SOFT, as described in Sec.~\ref{sec:synthdiagtheory}, the pitch angles of all particles increase as they move in to the high field-side of the device, and correspondingly they emit more radiation at shorter wavelengths. SYRUP, on the other hand, does not account for this effect; thus, the SYRUP spectra are shifted toward longer wavelengths compared to the corresponding \SOFT\ spectra. Furthermore, in certain scenarios where the spectrometer does not lie in the midplane and where particles with small pitch angles dominate emission, parts of the distribution function can be made completely invisible due to geometric effects, distorting spectra even further. Geometric effects are therefore essential and should always be taken into account when comparing simulations to experiments.


\section{Discussion and conclusions}
\label{sec:conclusions}
Synchrotron radiation is a powerful tool for diagnosing runaway electrons, but the analysis is complicated by the inhomogeneity of the tokamak magnetic field, the distribution of runaway electrons in phase-space and various geometric effects. It is therefore necessary to use a synthetic synchrotron diagnostic that takes into account the magnetic equilibrium and the properties of the detector used in experiments when studying the synchrotron radiation. \SOFT\ is a flexible tool that computes synchrotron images and spectra by solving the guiding-center equations of motion in arbitrary magnetic geometry and calculating the emitted radiation. It can easily be coupled with Fokker-Planck solvers to simulate physically meaningful scenarios.

The region of the runaway electron distribution function in momentum-space that contributes significantly to the detected synchrotron radiation is typically large, yielding a rather complex image. The image is however linear in the sense that it can be considered a weighted superposition of images resulting from various radially localized mono-energetic and mono-pitch populations of runaways, with the distribution of runaways acting as the weight function. This permits study of the effect of mono-energetic and mono-pitch distributions on the synchrotron spot, and conclusions from these studies to be applied to more complex cases involving runaways that are continuously distributed in momentum-space.

Since the amount of synchrotron radiation emitted by a particle is proportional to $p_\perp^2$ as indicated by~\eqref{eq:totalpower}, particles with large pitch angles tend to emit much more synchrotron radiation than those with small pitch angles, at a given energy. Conversely, particles with small pitch angles tend to be more numerous in the distribution function, and so the part of the distribution function that emits the most synchrotron radiation is a balance between these two properties. While a single particle mono-energy/mono-pitch angle spectrum or image can often be fitted to measurements, it is not useful in describing the actual distribution function, in which the fitted particle may only make up a vanishing fraction of the whole.

The synchrotron radiation spot shape and intensity distribution not only depends on the properties of the runaway electron population, but also on the placement of the synchrotron detector. This is an effect of the strong forward-beaming of synchrotron radiation, which must be emitted directly towards the detector for the particle to be seen. One of the more interesting consequences of this is that particles with small pitch angles, moving near the magnetic axis, will be invisible to the detector if the detector is placed too far from the midplane. While it may seem from this as though all synchrotron detectors should be placed in the midplane, it would be a good idea to distribute detectors both in, below and above the midplane, since this could further constrain the possible shape of the distribution function, which is the unknown quantity to be determined.

The great versatility of \SOFT~makes it possible to model experimental scenarios through forward modelling. This can be done with the use of a numerical kinetic-equation solver such as \textsc{CODE}~\cite{Landreman2014,Stahl2016}, \textsc{LUKE}~\cite{Peysson2003,Decker2004,Nilsson2015} or \textsc{NORSE}~\cite{Stahl2017}, which allow for experimental parameters such as density, temperature and electric field to be given as input. The forward modelling approach would also allow verification of kinetic theory for runaway electrons in a direct way.
Coupling \SOFT~with a numerical kinetic-equation solver to test whether any observed synchrotron images can be sufficiently constrained from measured quantities (that are of importance to runaway generation) is a useful way to model experiments.

While the framework developed in the beginning of this paper and leading up to Eq.~\eqref{eq:soft} was applied specifically to synchrotron radiation, there is nothing preventing it from being applied to other forms of radiation. Of particular interest when studying runaway electrons are hard X-rays from bremsstrahlung emission. Since bremsstrahlung is also emitted almost entirely in the particle's direction of motion, much of the theory presented in this paper is directly applicable to such studies and could be used to build a synthetic bremsstrahlung diagnostic.

\ack
The authors are grateful to G Papp and I Pusztai for fruitful discussions and the entire Alcator C-Mod team for excellent maintenance and operation of the tokamak. This work has been carried out within the framework of the EUROfusion Consortium and has received funding from the Euratom research and training programme 2014-2018 under grant agreement No 633053. The views and opinions expressed herein do not necessarily reflect those of the European Commission. The authors also acknowledge support from Vetenskapsr\aa det and the European Research Council (ERC-2014-CoG grant 647121). 

\bibliographystyle{iopart-num}
\bibliography{references}

\end{document}